%Paper: 9203060
%From: Free the RCFTs <MOORE%YALPH2.BITNET@yalevm.ycc.yale.edu>
%Date: Sat, 21 Mar 1992 21:49 EDT

\input harvmac.tex

% Poor man's Blackboard Bold characters often used :
\def\inbar{\,\vrule height1.5ex width.4pt depth0pt}
\def\IB{\relax{\rm I\kern-.18em B}}
\def\IC{\relax\hbox{$\inbar\kern-.3em{\rm C}$}}
\def\ID{\relax{\rm I\kern-.18em D}}
\def\IE{\relax{\rm I\kern-.18em E}}
\def\IF{\relax{\rm I\kern-.18em F}}
\def\IG{\relax\hbox{$\inbar\kern-.3em{\rm G}$}}
\def\IH{\relax{\rm I\kern-.18em H}}
\def\II{\relax{\rm I\kern-.18em I}}
\def\IK{\relax{\rm I\kern-.18em K}}
\def\IL{\relax{\rm I\kern-.18em L}}
\def\IM{\relax{\rm I\kern-.18em M}}
\def\IN{\relax{\rm I\kern-.18em N}}
\def\IO{\relax\hbox{$\inbar\kern-.3em{\rm O}$}}
\def\IP{\relax{\rm I\kern-.18em P}}
\def\IQ{\relax\hbox{$\inbar\kern-.3em{\rm Q}$}}
\def\IR{\relax{\rm I\kern-.18em R}}
\font\cmss=cmss10 \font\cmsss=cmss10 at 7pt
\def\IZ{\relax\ifmmode\mathchoice
{\hbox{\cmss Z\kern-.4em Z}}{\hbox{\cmss Z\kern-.4em Z}}
{\lower.9pt\hbox{\cmsss Z\kern-.4em Z}}
{\lower1.2pt\hbox{\cmsss Z\kern-.4em Z}}\else{\cmss Z\kern-.4em Z}\fi}
\def\IGa{\relax\hbox{${\rm I}\kern-.18em\Gamma$}}
\def\IPi{\relax\hbox{${\rm I}\kern-.18em\Pi$}}
\def\ITh{\relax\hbox{$\inbar\kern-.3em\Theta$}}
\def\IOm{\relax\hbox{$\inbar\kern-3.00pt\Omega$}}

\def\CR {{\cal R}}
\def\CD {{\cal D}}
\def\CF {{\cal F}}

\def\CO {{\cal O}}

\def\p {\partial}

\def\tb{\bar t}
\def\log {{\rm log}}

\Title{\vbox{\baselineskip12pt\hbox{YCTP-P7-92}
\hbox{hepth@xxx/9203060}}}
{\vbox{\centerline{Classical Scattering}
\centerline{in}
\centerline{1+1 Dimensional String Theory}}}

\centerline{Gregory Moore and Ronen Plesser}
\centerline{Department of Physics}
\centerline{Yale University}
\centerline{New Haven, CT 06511-8167}
\bigskip
\noindent
We find the general solution to Polchinski's classical
scattering equations for $1+1$ dimensional string theory.
This allows efficient computation of
scattering amplitudes in
the standard Liouville $\times$ $c=1$ background.
Moreover, the solution leads to a mapping
from a large class of
time-dependent collective field theory backgrounds
to corresponding nonlinear sigma models.
Finally, we derive recursion
relations between tachyon amplitudes. These may
be summarized by an infinite set of nonlinear
PDE's for the partition function in an arbitrary
time-dependent background.

\Date{March 18, 1992}
%\draft
\noblackbox

\newsec{Introduction}

Time-independent string backgrounds have been
much studied at the classical level. The time-independence
of the background allows the use of the methodology of
Euclidean 2D conformal field theory. On the other hand,
nontrivial time-dependent backgrounds ineluctably
lead to problems with negative signature bosons and negative
conformal weight vertex operators.
The corresponding conformal field theories
are expected to be more subtle than their Euclidean cousins.

While there are some isolated examples of time-dependent
backgrounds in string theory,
a large collection of tractable or solvable
backgrounds has only recently become available.
As a result of the recent developments in matrix model
technology
\nref\kleb{See, e.g., I.R. Klebanov ``String Theory in Two Dimensions,''
Princeton preprint PUPT-1271, hepth 9108019, for a review}%
\refs{\kleb}
and the parallel developments in Liouville theory
\nref\gervais{J.-L. Gervais, Phys. Lett. {\bf 243B}(1990)85;
Commun. Math. Phys. {\bf 130}(1990)257, and references
therein.}%
\nref\natiliouv{N. Seiberg, ``Notes on Quantum liouville Theory and
Quantum Gravity,'' in
{\it Common Trends in
Mathematics and Quantum Field Theories,} Proceedings of the 1990 Yukawa
International Seminar, Prog. Theor. Phys. Supp. {\bf 102}.}%
\nref\joetalk{
J. Polchinski, ``Remarks on the Liouville Field Theory,''
UTTG-19-90, to appear in Strings '90, Texas AM.}%
\nref\kutrev{D. Kutasov, ``Some Properties of (Non) Critical
Strings,'' Princeton preprint PUPT-1277, hepth 9110041}%
\nref\martinec{E. Martinec, ``An Introduction to 2d Gravity and Solvable
String Models,'' Rutgers preprint RU-91-51, hepth 9112019.}%
\refs{\gervais {--}\martinec},
we now have available for study a tractable but nontrivial
example of string theory, namely, string theory in
$1+1$ dimensions. Using the collective
field theory obtained from the matrix model
\nref\dj{S.R. Das and A. Jevicki, Mod. Phys. Lett. {\bf A5} (1990) 1639}%
\nref\senwad{A.M. Sengupta and S.R. Wadia,
``Excitations and interactions in $d=1$ string theory,''
Int. Jour. Mod. Phys. {\bf A6} (1991) 1961.}%
\refs{\dj,\senwad},
Polchinski found a large class of nontrivial
time-dependent classical backgrounds for 1+1 dimensional
string theory
\nref\joesea{J. Polchinski, ``Classical limit of 1+1 Dimensional
String Theory,'' Nucl. Phys. {\bf B362}(1991)125.}%
\refs{\joesea}.
Unfortunately, the physics of these solutions has been
partly obscure since the relation of the matrix model
coordinate $\tau$
\refs{\dj,\senwad}
to the Liouville coordinate $\phi$ is nontrivial
\ref\msi{G. Moore and N. Seiberg, ``From Loops to Fields in
2D Gravity,'' to appear in Int. Jour. Mod. Phys.}.
Thus the mapping of the solutions of \joesea\ to
a corresponding nonlinear $\sigma$-model has been
unclear, even at the formal level.

In this note we investigate further Polchinski's scattering equations
(equations (2.5) below), and give the general solution of
these equations in (3.4). From the classical solution
we can extract genus zero $S$-matrix elements. For example,
the result (4.4) extends the result of
\ref\kdf{D. Kutasov and Ph. DiFrancesco, Phys. Lett. {\bf 261B}(1991)385;
D. Kutasov and Ph. DiFrancesco, ``World Sheet and Space Time
Physics in Two Dimensional (Super) String Theory,''
Princeton preprint PUPT-1276, hepth 9109005.}\
for $1\to m$ scattering to
$n \to m$ scattering in a simple kinematic regime.
We then propose
a (formal) mapping from a given solution of Polchinski's
equations to a nonlinear sigma model in (5.8).
Using the solution (3.4) we derive some recursion relations
between tachyon amplitudes which are summarized by the
nonlinear differential equations (6.6) for the partition function.
We conclude with hopes for the future.

\newsec{Classical Solutions of 2D String Theory}

Classical 2D string theory can be formulated in terms
of a single field theoretic degree of freedom, $\chi(\lambda,t)$
related to the ``massless tachyon'' degree of freedom.
The (collective field theory)
Hamiltonian is
\nref\gki{D. Gross and I. Klebanov,
``Fermionic String Field Theory of $c=1$ 2D Quantum Gravity,''
Nucl. Phys. {\bf B352} (1991) 671.}%
\refs{\dj,\senwad,\gki}
\eqn\djham{
H=\int d\lambda \biggl\{ {g^2\over 2}\chi' \pi_\chi^2  + {\pi^2\over 6 g^2}
(\chi')^3+{v(\lambda)\over g^2}\chi'\biggr\}
}
where $v(\lambda)$ is the double-scaled matrix model potential.
Physically appropriate boundary conditions for the fields have
not been carefully investigated. Roughly speaking, we require
$\pi_\chi,\chi$ to vanish on a half-axis. Moreover
$\p_\lambda \chi\geq 0$.
The classical field equations are nonlinear and look
formidable, but
by defining $p_\pm \equiv - g^2 \pi_\chi \pm \pi \chi'$ the equations
separate
\nref\jevi{J. Avan and A. Jevicki,
``Classical Integrability and Higher Symmetries of Collective Field
Theory,'' Brown preprint BROWN-HET-801.}%
\nref\gkii{D.J. Gross and I.R. Klebanov, ``$S = 1$ for $c = 1$,''
Nucl. Phys. {\bf B359} (1991) 3.}%
\refs{\senwad,\joesea,\jevi,\gkii}
\eqn\sepeqs{
\p_t p_\pm=-v'(\lambda)-p_\pm \p_\lambda p_\pm\ .
}

In particular, for the potential $v(\lambda)=-\half \lambda^2$
the general solution to these equations was found by
Polchinski \joesea. He also gave a beautiful interpretation
in terms of a time-dependent Fermi sea in free fermion
phase space with coordinates $(\lambda,p)$, determined by the
parametric equations:
\eqna\frms
$$
\eqalignno{
\lambda &=(1+a(\sigma))\cosh(\sigma-t)&\frms a\cr
p&=(1+a(\sigma))\sinh(\sigma-t)\ .& \frms b\cr}$$

For $a(\sigma)$ sufficiently small and sufficiently slowly-varying
(see below)
\frms{a}\ will have two solutions $\sigma_\pm(\lambda,t)$, and
substitution into \frms{b}\ defines upper and
lower branches $p_\pm(\lambda,t)$ of the sea.
Using the change of variables $\lambda=\cosh \tau$, $\tau>0$,
we define the
asymptotic waveforms $\psi_\pm(t\pm \tau)$ in the far past
and future by the limiting behavior
\eqn\limbeh{
p_\mp(\lambda,t)\rightarrow \mp \lambda \pm {1\over 2\lambda}\bigl(1+
\psi_\pm(t\pm \tau)\bigr)+\CO(1/\lambda^2)
}
where $\lambda\to +\infty$ holding $t\pm \tau$ fixed.

The essential remark \joesea\ is that the incoming and outgoing
asymptotic waveforms are related by a diffeomorphism
$x\mapsto \tilde x$, determined by
the classical solution: $\tilde x=x+\log(1+\psi_-(x))$.
Thus, the basic equation of scattering, by
which we can determine the outgoing waveform $\psi_-$
in terms of the incoming waveform $\psi_+$ is the
functional equation
\eqn\scateq{
\eqalign{
\psi_-(x)&=\psi_+(\tilde x)\cr
&=\psi_+\bigl( x+\log(1+\psi_-(x))\bigr) \ .\cr}
}
One can study the time-reversed process using
the inverse diffeomorphism, i.e.,  $\psi_+(y)=\psi_-(\check y)$,
where $\check y=y-\log(1+\psi_+(y))$. The asymptotic
waveform $\psi_\pm(x)$ completely determines the Fermi sea
profile through the relation
\eqn\detprof{
a\bigl({x+\tilde x \over 2}\bigr)=\sqrt{1+\psi_-(x)}-1\ .
}
Thus, the space of solutions of the equations of motion
may be formally identified with the diffeomorphisms of the line.

We end this section with a few comments on the range of validity of
\scateq. First note that \detprof\ makes sense only for $\psi_{\pm}\ge -1$.
Indeed, we can see directly from \limbeh\ that
violating this condition yields
solutions in which the Fermi sea extends beyond the quadrant
$\lambda > |p|$; these represent other phases of the theory
in which the limiting behavior differs from \limbeh. If this constraint is
met we find $a > -1$,
hence from \frms{}\ $p_+(x,t) = p_-(x,t)$ has a
unique solution (at the classical turning point $p_{\pm} = 0$) so
$\p_\lambda \chi =p_+-p_-\ge 0$.
The second comment is that if $\psi_-$ satisfies the restriction above,
the map $x \mapsto\tilde x$ is a diffeomorphism provided
\eqn\consis{
1 + {\psi_-' \over 1 + \psi_-} \ge 0\ .
}
When this condition is violated, \frms{a}\
will have more than two solutions for large enough $t$,
representing a ``fold'' in the Fermi
sea (a region where there are four or more branches of the sea
at fixed $\lambda$). In terms of the original
collective field theory, this corresponds to a singularity
in the field $\chi$.
The fact that the classical field equations
evolve smooth initial data to singular field
configurations is reminiscent of singularities
occurring in nonlinear field theories, notably, in
general relativity. We comment further on this in
the conclusions.

\newsec{General Solution to the Classical Scattering
Equations}

Suppose  $\Psi_\pm$ constitute a solution of
the classical scattering equations \scateq , and suppose
further that
$\Psi_\pm + \gamma_\pm$ is a nearby solution,
where $\gamma_\pm$ are small. To first order in
the variations \scateq\ becomes
\eqn\perti{
\gamma_+(\tilde x)d\tilde x=\gamma_-(x) dx
}
where $\tilde x=x+\log(1+\Psi_-(x))$.
Taking a Fourier transform of this equation, with
\eqn\dffour{
\gamma_\pm(x)\equiv \int_{-\infty}^\infty  \gamma_\pm(\xi)
e^{i\xi x}d\xi
}
leads to
\eqn\solfour{
\gamma_\pm(\xi)={1\over 2\pi}
\int e^{-i \xi x}\gamma_\mp(x)(1+\Psi_\mp(x))^{\mp i \xi} dx\ .
}
This may be regarded as a first-order differential equation in
function space. Integrating this equation with the
boundary condition $\psi_+=0 \Rightarrow \psi_-=0$
we have the general
solution of Polchinski's scattering equations:
\eqn\gensol{
2\pi \psi_\pm(\xi)={1\over 1\mp i\xi}
\int_{-\infty}^\infty e^{-i\xi x}
\biggl[(1+ \psi_\mp(x))^{1\mp i\xi}-1\biggr]dx\ .
}
or, in position space:
\eqn\gensoli{
\psi_{\pm}(x)=-\sum_{p\geq 1}{\Gamma(\pm\p_x+p-1)\over \Gamma(\pm\p_x)}
{(-\psi_\mp(x))^p\over p!}
}

The solution \gensol\ is valid for finite (i.e., not infinitesimal)
field configurations. The formula breaks down exactly
for solutions violating the conditions of the previous section.
For example,
a waveform $\psi_-(x)=\beta \cos \omega_0 x$ in the far future
corresponds to the incoming waveform:
\eqn\inwvfrm{
\eqalign{
\psi_+(x)=&Re\biggl(\sum_{n\geq 0} \psi_+(n) e^{i n \omega_0 x}\biggr)\cr
\psi_+(n)=&2 \pi (-1)^n {\Gamma(i n \omega_0
+n -1)\over \Gamma(i n \omega_0) n!}\bigl({\beta\over
2}\bigr)^n {}_2F_1({i n \omega_0+n-1\over 2},
{i n \omega_0+n\over 2};n+1;\beta^2)\ .\cr}
}
The hypergeometric function has a
branch point at $\beta=1$ reflecting the breakdown of
\detprof. Now suppose $\beta<1$ and
consider the dependence on
$\omega_0$. Using stationary phase approximation in
\gensol\ we find that for
\eqn\foldcond{
{\beta\omega_0\over \sqrt{1-\beta^2}}<1
}
$\psi_+(n)$ decreases exponentially with $|n|$
while if condition \foldcond\ is violated then
$\psi_+(n)\sim \CO(n^{-3/2})$ for large $n$.
Thus the Fourier series $\psi_+(x)$ becomes
discontinuous, reflecting the violation of \consis\ (in this
periodic solution there are in fact an infinite number of ``folds'').

A final interesting pathology occurs for waves formally
corresponding to $\omega=i$. In this case the
functional equations \scateq\ can be solved directly to
give the solution:
\eqn\disassol{
\eqalign{
\pi_\chi&=-{\alpha e^t\over 2 g^2} ~\theta(\lambda\geq 1+\half \alpha e^t)\cr
\p_\lambda \chi&={1\over \pi}\sqrt{(\lambda-1-\half \alpha e^t)
(\lambda+1-\half \alpha e^t)}~\theta(\lambda\geq 1+\half \alpha e^t)
\ .\cr}
}
In the infinite past we have the standard static background,
but as time increases the Fermi sea is drained, leaving nothing
behind. This pathology can be eliminated by the boundary
condition $\p_\lambda \chi=0\Rightarrow \pi_\chi=0$
forcing zero momentum flow at the edge of the sea. While
physically reasonable, it is not clear that this eliminates all
pathological solutions.

\newsec{Tree-Level S-Matrix in a Classical Background}

\subsec{Scattering in the standard background}

We now introduce quantum mechanics by considering the
$\gamma_\pm$ to be related to incoming and outgoing
free quantum fields via:
\eqn\defflds{\eqalign{
\gamma_\pm&\rightarrow \sqrt{4 \pi} g
(\p_t\ \pm  \p_\tau) \chi_\pm\cr
\chi_+&=i\int_{-\infty}^\infty \alpha_+(\xi)
e^{i \xi (t+\tau)} {d\xi\over \sqrt{4 \pi} \xi}\cr
\chi_-&=i\int_{-\infty}^\infty \alpha_-(\xi)
e^{i \xi (t-\tau)} {d\xi\over \sqrt{4 \pi} \xi}\cr
[\alpha_\pm(\xi),\alpha_\pm(\xi')]&= - \xi \delta(\xi+\xi')\ .\cr}
}

The general solution \gensol\ of the classical scattering
equations implicitly summarizes the entire tree-level
$S$-matrix. Following Polchinski, we
interpret the expansion of \gensol\ around the trivial
background as the quantum mechanical operator equation
relating in and out fields:
\eqn\inout{
\alpha_\pm(\eta)= \sum_{p\geq 1} (-2g)^{p-1}
{\Gamma(1\mp i\eta)\over \Gamma(2\mp i\eta-p)}
{1\over p!}\int_{-\infty}^\infty d^p\xi
\delta(\eta-\sum \xi_i) : \alpha_\mp(\xi_1)\cdots
 \alpha_\mp(\xi_p):\ .
}
The vacuum is defined by $\alpha_\pm(-\omega)|0\rangle=0$.
(In this paper
$\omega$ will always stand for {\it positive} quantities.)
The expression on the rhs of \inout\ is normal-ordered
with respect to this vacuum.%
\foot{Actually, the normal ordering in \inout\
is a convenient choice;
to lowest order in $g$ the operator ordering is irrelevant.}

Incoming (outgoing) states are created by the action of
$\alpha_+(\omega)$ ($ \alpha_-(\omega)$).
We compute the tree level approximation to the connected
$S$-matrix element
\eqn\selmnt{
S_c(\sum \omega_i\to \sum \omega_i') =
\langle 0|\prod_i  \alpha_-(-\omega'_i)
\prod_j \alpha_+(\omega_j) |0\rangle_c
}
by inserting \inout\ and extracting the connected contribution at
lowest nonvanishing order in $g$.
The connected amplitude for $n\to m$ scattering is
of order $g^{m+n-2}$, and calculations of
tree-level scattering amplitudes reduce to combinatorics.
For example, consider
$S(\sum_1^n\omega_i \to \sum_1^m\omega'_i)$ in the
kinematic regime $\forall k,
\omega_n > \omega'_k > \sum_{j=1}^{n-1} \omega_j$.
Using \inout\ one finds the simple expression
\eqn\ntom{
S = -i (-2g)^{m+n-2} \prod_{j=1}^n \omega_j
\prod_{k=1}^m \omega'_k {\Gamma(-i\omega_n) \over \Gamma(1-m-i\omega_n)}
{\Gamma(1-m-i\Omega) \over \Gamma(3-n-m-i\Omega)}
}
where $\Omega = \sum_{j=1}^n \omega_j$.
Setting $n=1$, we recover the amplitudes computed in \kdf\ using a continuum
formulation.

\subsec{Scattering in nontrivial backgrounds}

The formalism described above also allows us to compute
tree level quantum scattering amplitudes in nontrivial backgrounds.
There are two physical effects. The classical background $\psi_\pm$
can emit and absorb quanta, and the quanta themselves
can interact. The description of these two processes
must be self-consistent. We will assume the existence of
a classical detector which can exist in the background
and measure the presence of individual quanta
\ref\birrdav{This is not a trivial assumption.
See N.D. Birrell and P.C.W. Davies, {\it Quantum fields in
curved space}, Cambridge 1982, for a discussion of
some of the issues involved.}.

The effects of the classical background
are taken into account by writing the total in- and
out- fields as a classical piece plus a
quantum piece: $\hat \psi_\pm = \psi_\pm + \p_\tau \chi_\pm$.
Thus the total field $\hat\psi_+$ is related to
the quantum field by a unitary transformation
\eqn\untrm{
S[\psi_+]\equiv e^{-{1\over 2 g}
\int^\infty_{-\infty} {d\xi\over\xi} \alpha_+(\xi)\psi_+(-\xi)}
}
such that $S^{\dagger}[\psi_+] \gamma_+(\xi) S[\psi_+] =
\gamma_+(\xi) + \psi_+(\xi)$.
An analogous transformation  $S[\psi_-]$ shifts $\gamma_-$.
The effect of this unitary transformation on $\hat \psi_{\mp}$ is quite
nontrivial and leads to an equation generalizing \inout.
A classically evolving detector measures
the $S$-matrix given by
\eqn\smatrix{
S(\sum \omega_i \to \sum \omega '_i;\psi) =
\langle 0|S^\dagger[\psi_-]
\prod \alpha_-(-\omega_i)\prod \alpha_+(\omega_i)
S[\psi_+]|0\rangle
}
which may be computed using the above formulae.

It is also of interest to consider particle creation
in the state which is described as an initial classical
state with no ``extra'' quanta $|\psi_+\rangle\equiv
S[\psi_+]|0\rangle$. For example we find
\eqn\prtcrt{
\eqalign{
\langle \psi_+ | \alpha_-(\omega) &\alpha_-(-\omega) |\psi_+\rangle
 = \cr
&\int_0^\infty \! \eta d\eta \,
\Bigl| \sum_{p=2}^\infty
{\Gamma(1+i\omega)\over\Gamma(2+i\omega-p)(p-1)!}\int\! d^{p-1}\xi\,
\delta(\omega+\eta-\sum_1^{p-1} \xi_i) \prod_1^{p-1} \psi_+(\xi_i)
\Bigr|^2\cr}
}
to lowest order in $g$ and all orders in the background.

\newsec{Vertex Operator Calculations}

We would like to describe the scattering in
time-dependent backgrounds in terms of vertex
operator correlators in time-dependent
$\sigma$-model backgrounds. First, we relate
the above $S$-matrix amplitudes to the vertex
operator correlators of the
Euclidean $c=1$ $\times$ Liouville
system with action:%
\foot{We follow the notation of \natiliouv.}
\eqn\action{
S=\int d^2z \sqrt{\hat g}\biggl[{1\over 8 \pi}
(\hat\nabla \phi)^2
+{\mu\over 8 \pi \gamma^2}e^{\gamma \phi}+{Q\over 8 \pi}
\phi R(\hat g)+{1\over 8 \pi} (\hat\nabla X)^2
\biggr]}

Vertex operator correlators for the
theory \action\ have been calculated
in \kdf. The gravitationally dressed
vertex operators are
\eqn\eucvert{\eqalign{
V_q&=\int e^{i q X/\sqrt{2}} e^{\sqrt{2}(1-\half|q|)\phi}\cr
\bar V_q&=\int e^{i q X/\sqrt{2}} e^{\sqrt{2}(1+\half|q|)\phi}\cr}
}
where $q\in \IR$. The operators $\bar V_q$ are known
to be subtle. It is argued in \natiliouv\ that
they either do not exist as quantum operators or that
they decouple from ordinary amplitudes. The correlators
of the $V_q$'s may be analytically continued to
obtain the S-matrix elements of section (4.1) above.
Specifically, the amplitudes
\eqn\eucamp{
\CR(q_i;q_i')\equiv
\langle \prod_{i=1}^n V_{q_i} \prod_{i=1}^m V_{q_i'}\rangle
}
where $q_i<0$, $q_i'>0$, and $\sum q_i+\sum q_i'=0$
are related to the connected
Minkowskian $S$-matrix elements by the analytic continuation
\nref\mpr{G. Moore, R. Plesser, S. Ramgoolam, ``Exact S-Matrix for
2D String Theory,'' Yale preprint P35-91, hepth 9111035.}%
\refs{\kleb,\mpr}
\eqn\ancon{
\CR(q_i=i\omega_i;q_i'=-i\omega_i')=
\prod_{i=1}^n {\Gamma(i\omega_i)\over \Gamma(-i\omega_i)}
\prod_{i=1}^m {\Gamma(i\omega_i)\over \Gamma(-i\omega_i)}
S_c(\sum_{i=1}^n\omega_i\to \sum_{i=1}^m\omega_i')\ .
}
Since the sign of $q$ is crucial to the Minkowskian
interpretation we must distinguish four different
Minkowskian vertex operators:
\eqn\minkvert{
\eqalign{
T^+_{-\omega}&\equiv{\Gamma(-i\omega)\over\Gamma(i\omega)}\int_\Sigma
e^{2\varphi} e^{i \omega (t+\varphi)} \cr
T^-_\omega&\equiv{\Gamma(-i\omega)\over\Gamma(i\omega)}\int_\Sigma
e^{2\varphi} e^{-i \omega (t-\varphi)} \cr
T^+_{\omega}&\equiv{\Gamma(i\omega)\over\Gamma(-i\omega)}\int_\Sigma
e^{2\varphi} e^{-i \omega (t+\varphi)} \cr
T^-_{-\omega}&\equiv{\Gamma(i\omega)\over\Gamma(-i\omega)}\int_\Sigma
e^{2\varphi} e^{i \omega (t-\varphi)} \cr}
}
where $\varphi = \phi/\sqrt{2}$.
We see that $T^+_{-\omega},T^-_{\omega}$ are the continuation
of the Seiberg branch $V_q$, while $T^+_{\omega},T^-_{-\omega}$
are the continuation of the other branch $\bar V_q$.
Thus, in Minkowski space we see that for scattering
on the left half-line the Seiberg branch is simply the branch for which
the incoming (outgoing) particles are right- (left-) moving.
Assuming that the Minkowskian amplitudes are correlation functions of
analytically continued vertex operators,
we identify the vertex operators creating incoming and outgoing tachyons
in Minkowski space as
\eqn\identify{\eqalign{
V_{in}(\omega)=& \, T^-_{\omega}\cr
V_{out}(\omega)=& \, T^+_{-\omega}\ .\cr}
}

Armed with this identification we can now write \smatrix\ as
a $\sigma$-model correlator. We first note that
$S[\psi_{\pm}]$ may be normal ordered to give
\eqn\snorord{
S[\psi_{\pm}] = e^{-{1\over 4g^2}
\int_0^\infty {d\xi\over\xi} |\psi_{\pm}(\xi)|^2}\
e^{-{1\over 2g} \int^\infty_0 {d\xi\over\xi}\,
\alpha_{\pm}(\xi)\psi_{\pm}(-\xi)}\
e^{-{1\over 2g} \int^0_{-\infty} {d\xi\over\xi}\,
\alpha_{\pm}(\xi)\psi_{\pm}(-\xi)}\ .
}
Inserting this converts \smatrix\ to a sum of correlation functions of the
form \ancon, and we may thus write
\eqn\smbak{\eqalign{
\langle \psi_-|
\prod \alpha_-(-\omega_i) &\prod
\alpha_+(\omega_i') |\psi_+\rangle =
e^{-{1\over 4g^2} \int_0^\infty {d\xi\over\xi}
(|\psi_+(\xi)|^2-|\psi_-(\xi)|^2)} \cr
&\langle \prod V_{out}(\omega_i) \prod V_{in}(\omega_i')
e^{-{1\over 2g}\int_0^\infty {d\omega\over\omega}
(\psi_-(\omega) V_{out}(\omega) + \psi_+(-\omega) V_{in}(\omega))}
\rangle_{\sigma-{\rm model}}\cr}
}
realizing the time-dependent tachyon background explicitly as a
modification of the $\sigma$-model action.

We close with some comments on the identification \identify\ of the vertex
operators. The first remark is that, if we assume that the analytically
continued $\bar V_q$ decouple as in Euclidean space, there is an ambiguity
in the identification because mixing with these will not affect amplitudes.
This might resolve a puzzle regarding the
relation of the vertex operators to
the asymptotics of the the Wheeler-de-Witt
wavefunction
\nref\mss{G. Moore, N. Seiberg, and M. Staudacher, ``From Loops to
States in Two-Dimensional Quantum Gravity,''
Nucl. Phys. {\bf B362}(1991)665}%
\refs{\mss,\msi}.%
\foot{We thank N. Seiberg for lively correspondence on this
and related questions.}
A related point is that if one attempts to construct a $\sigma$-model
formulation of amplitudes like \prtcrt\ one is forced to include the
``wrong'' branch operators and abandon the hypothesis of decoupling.

\newsec{Relations for tachyon amplitudes}

\subsec{Recursion relations}

As a final application of \inout, we derive
some ``Ward identities'' for insertion of a tachyon operator
in Euclidean vertex operator correlators.
In this section we scale $-2g$ to one for
simplicity.

The identities are most simply written in terms of
$\CT_q = {\Gamma (|q|) \over \Gamma (1-|q|)} V_q$.
Consider first the insertion of a ``special tachyon,''
with $q\in \IZ_+$.
If we continue $\omega\to i n$ with $n\in \IZ_+$ then
the series \inout\ truncates after $n+1$ terms. These
terms have a ``universal'' effect in correlation functions.
\eqnn\ward %
Specifically, an insertion of $\CT_n$ is given by%
\foot{The case $m_-=m$ is exceptional
(\ward\ vanishes while the correlator does not) but
the amplitude is known
from \ntom. This ungainly feature will be remedied below.}
$$\eqalignno{
\langle \CT_n &\prod_{i=1}^m \CT_{q_i} \rangle
= \sum_{k=2}^{\min (m,n+1)}
{\Gamma(n) \over \Gamma(2+n-k)} &\ward\cr
&\sum_{l=1}^{\min (m_-,k-1)} \sum_{|T|=l} \theta(-q(T)-n)
\sum_{S_1, \ldots S_{k-l}}
\prod_{j=1}^{k-l} \Bigl[ \theta(q(S_j)) q(S_j)
\langle \CT_{-q(S_j)} \prod_{S_j} \CT_{q_i} \rangle \Bigr] \cr
}$$
The notation is the following: Let $S = \{q_1 \ldots q_m\}$,
and let $S^-$ denote the subset of $S$ of negative momenta.
Denote $m_-=|S_-|$.
The sum on $T$ is over subsets of $S^-$ of order $l$.
The subsequent sum is over distinct disjoint decompositions
$S_1 \amalg \ldots  \amalg S_{k-l}= S\setminus T$. $q(T)$
denotes the sum of momenta in the set $T$.
The momenta $q_i$ are taken to be generic so
that the step functions are unambiguous. This entails no
loss of generality since
the amplitudes are continuous (but not differentiable)
across kinematic boundaries.
\nref\dsft{G. Moore, ``Double Scaled Field Theory at $c=1$,''
Nucl. Phys. {\bf B368} (1992) 557.}%

The first two examples of \ward\ are:
\hfill
\break
$n=1$:
\eqn\bndryop{
\langle  \CT_{1}\prod_{i=1}^n\CT_{q_i}\rangle=
 \sum_{q_i<-1}
|q_i+1| \langle \CT_{q_i+1}\prod_{j\not= i}\CT_{q_j}\rangle
}
$n=2$:
\eqn\wardtwo{\eqalign{
\langle &\CT_2 \prod_{i=1}^m \CT_{q_i} \rangle
=  \sum_{q_i < -2} |q_i + 2|
\langle \CT_{q_i+2} \prod_{j\ne i} \CT_{q_j} \rangle \cr
&+
\sum_{{\scriptstyle q_i+q_j < -2}\atop{\scriptstyle q_i,q_j<0}}
|q_i+q_j+2|
\langle \CT_{q_i+q_j+2} \prod_{k\ne i,j} \CT_{q_k} \rangle \cr
& + \sum_{q_i<-2} \sum_{S_1\amalg S_2=S\setminus \{ q_i\} }
\theta(q(S_1))q(S_1)\theta(q(S_2))q(S_2)
\langle \CT_{-q(S_1)} \prod_{S_1} \CT_{q_j} \rangle
\langle \CT_{-q(S_2)} \prod_{S_2} \CT_{q_j} \rangle \ .\cr}
}
Note that in \wardtwo\ there is a change in tachyon number
by one in the second line,
and the product of two correlators in the third line. The pattern
continues for higher $n$: there are terms with $|T|=l=k-1$ removing
$l$ incoming tachyons, which are linear in the correlators, and
terms with a product of $k-l$ correlators.

A little reflection shows that it is easy to generalize the
identity \ward\ to the case of an insertion of
$\CT_p$ with $p\in\IR$ merely by allowing the sum on
$k$ to run to $m$. The resulting tachyon recursion relations
give an efficient way to calculate amplitudes
since applying \ward\ to the tachyon of largest absolute
momentum reduces the number of tachyon insertions by at
least two.
For example, the reader may quickly calculate the
five point function in an arbitrary kinematic configuration.

\subsec{Differential equations}

The identities \ward\ are elegantly expressed in terms of the generating
functional for connected amplitudes
\eqn\genfnl{
\CF\bigl[ t,\tb\bigr] =
\langle 0| e^{\int_0^\infty (t(p) \CT_p + \tb(p) \CT_{-p}) dp}
|0\rangle_c\ .
}
Introducing the nonlinear differential operator%
\foot{For $b=0$ (corresponding to the exceptional case mentioned below \ward)
the last product is taken to be one.}
\eqn\ewardw{
\eqalign{
\bar \CD_n \CF \equiv \sum_{a=1}^n \sum_{b=0}^{n+1-a}
{(n-1)!\over a! b! (n+1-a-b)!}\cr
\int_0^\infty \prod_1^a dp'_i \prod_1^b dp_j &\delta(\sum p'_i - \sum p_j - n)
\prod_1^a \tb(p'_i)
\prod_1^b \bigl( p_j {\delta \CF \over \delta \tb(p_j)} \bigr)\ .\cr}
}
and its conjugate $\CD_n$ we may summarize \ward\ as the
set of differential equations:
\eqn\diffeqs{\eqalign{
{\delta\CF\over \delta t(n)}&=\bar \CD_n \CF\cr
{\delta\CF\over \delta \bar t(n)}&=\CD_n \CF\ .\cr}
}
By (Euclidean) time reversal invariance $\CF(t,\bar t)=\CF(\bar t,t)$,
so only half the equations in \diffeqs\ are independent.

The equations \diffeqs, being equivalent
to recursion relations for tachyon amplitudes, uniquely determine
$\CF$, at least for
small $t,\bar t$.  Although there are $\aleph_0$ equations in $\aleph_1$
variables the correlation functions of $\CT_p$ are polynomials
in $p$, and hence completely specified by their values for
$p\in \IZ$. Thus one may restrict attention to $t(p),\bar t(p)$
with support on $\IZ_+$. Alternatively, one can write a larger set of
differential equations by generalizing the operators to
$\CD_p$ for $p\in\IR$. The essential difference is
that the sum on $a,b$ becomes infinite; in practice,
in the calculation
of a given amplitude the series terminates.

The consistency conditions obtained by equating mixed partial
derivatives in \diffeqs\ are complicated. We may write
\eqn\vir{
\bar \CD_n=\int_0^\infty dp \bar t(p+n) p
{\delta\over \delta \bar t(p)}+\cdots}
Thus the ``first'' term ($a=b=1$) in \ewardw\ defines a set of
operators satisfying half a Virasoro algebra, but the higher terms
introduce essential complications. It would be interesting to find a
concise algebraic description%
\foot{undoubtedly related to $W_{\infty}$}
of the algebra of operators generated by the equations
\diffeqs.

\subsec{Relation to other work}

Ward identities related to special states and special
tachyons have been the subject of much study over the
past year
\nref\wvii{D. Minic, J. Polchinski, and Z. Yang, ``Translation-Invariant
Backgrounds in 1+1 Dimensional String Theory,''
Texas preprint UTTG-16-91.}%
\nref\wviii{J. Avan and A. Jevicki,
``Classical Integrability and Higher Symmetries of Collective Field
Theory,'' Brown preprint BROWN-HET-801;
``Quantum Integrability and Exact Eigenstates of the Collective String
Field Theory,'' BROWN-HET-824.}%
\nref\wix{S.R. Das, A. Dhar, G. Mandal, S. R. Wadia,
``Gauge Theory Formulation of the c=1 Matrix Model: Symmetries and Discrete
States,'' IAS preprint IASSNS-HEP-91/52.}%
\nref\wx{E. Witten, ``Ground Ring of Two Dimensional String Theory,''
IAS preprint IASSNS-HEP-91/51.}%
\nref\polyakov{I.R. Klebanov and A.M. Polyakov, ``Interaction of
Discrete States in Two-Dimensional String Theory,'' hepth 9109032}
\nref\kms{D. Kutasov, E. Martinec, and N.
Seiberg, ``Ground Rings and Their Modules
in 2D Gravity with $c\le 1$ Matter,'' Princeton preprint
PUPT-1293; Rutgers preprint RU-91-49, hepth 9111048}%
\nref\kit{Y. Kitazawa, ``Puncture Operators in $c=1$ Liouville
Gravity,''  hepth 9112021}%
\nref\klebi{I. Klebanov,``Ward identities in two-dimensional
string theory,'' Princeton preprint, hepth 9201005}%
\nref\tanii{Yoichiro Matsumura, Norisuke Sakai, Yoshiaki Tanii,
``Interaction of Tachyons and Discrete States in $c \! = \! 1$ 2-D Quantum
Gravity,'' hepth 9201066}%
\nref\witzwie{E. Witten and B. Zwiebach, ``Algebraic Structures
and Differential Geometry in 2D String Theory,'' IASSNS-HEP-92/4,
hepth 9201056.}%
\nref\erik{E. Verlinde, ``The master equation of 2D string theory,''
IAS preprint IASSNS-HEP-92/5, hepth 9202021.}%
\refs{\msi,\wvii{--}\erik}.
We will now discuss the relation of \diffeqs\ to
some of these works.

In the formulation of
\refs{\msi,\dsft} one extracts the correlators of tachyons
as the coefficients of nonanalytic powers
$\ell^{|p|}$ in the small $\ell$ expansion of macroscopic
loop amplitudes.
The macroscopic loop operator carrying momentum $p$
has an expansion as a sum of local operators \msi
\eqn\locopexp{W_{in}(\ell,p)= -\CT_p {\pi|p|\over sin\pi |p|} \mu^{-|p|/2}
I_{|p|}(2\sqrt{\mu}\ell)-
\sum_{r=1}^\infty\hat\CB_{r,p}{2(-1)^r r\over r^2-p^2}
\mu^{-r/2}I_r(2\sqrt{\mu}\ell) \ .}
The operators $\hat\CB_{r,p}$ are redundant for $p\notin \IZ$,
and obey Ward identities organized by $W_{1+\infty}$ \msi.
Examining \locopexp, we see that the smoothness of
macroscopic loop amplitudes at $p\in \IZ$ implies the
identification $\CT_{\pm n }= \hat\CB_{n,\pm n}$ in correlators.
The identities we
find are thus related to the boundary operator Ward identities of
\msi. For example, taking the $\ell\to 0$ limit
of the boundary operator ($\CB_1$) Ward identity of \msi\
we obtain \bndryop.%
\foot{Reference \kit\ argued that this identity is
analogous to the puncture operator equation
at $c<1$.}
In general,
\ward\ is not identical to the Ward identities of \msi. A
detailed examination shows that the added terms in \ward\ are
absent in the cases which were explicitly
proved in \msi\ by manipulation of macroscopic loop amplitudes.
Evidently, in the general case there are complications in the
behavior of the measure for the fermion path integral
under $W_\infty$ transformations.

Other works have made direct use of the continuum Liouville
formulation. In \kms\ the method of bulk amplitudes is
applied to obtain linear relations on correlation functions
at finite string coupling.
As noted in \kms\ it is difficult to control the contact
terms which correct these relations so we cannot compare.

More recent studies, e.g., \refs{\klebi,\witzwie,\erik},
have examined the
theory at $-2g={1\over \mu}=\infty$ so it is not clear we should compare
with results at $-2g={1\over \mu}=1$. Nevertheless, we will try.
In \ward\ the term $l=k-1$ corresponds to a process
where $l$ (incoming) tachyons in the set
$T$ have been eliminated from the correlation function. This is
strongly reminiscent of the fact that  $W_\infty$ symmetry
generators do not preserve tachyon number.%
\foot{
This nonlinearity has been noted in \refs{\mpr,\kms,\klebi,\witzwie}.}
The Ward identities of \refs{\witzwie,\erik} constitute a set of beautiful
quadratic relations on amplitudes with insertions of all BRST invariant
dimension zero operators. In addition to the dressed tachyons
of ghost number two there are other BRST invariant operators
\nref\lz{B.Lian and G. Zuckerman, ``New Selection Rules and
Physical States in 2D Gravity,'' Phys. Lett.
{\bf B254} (1991)417;Phys. Lett. {\bf B266}(1991) 21}%
\nref\mukhi{S. Mukherji, S. Mukhi, and A. Sen, Phys. Lett.
{\bf B266}(1991)337}%
\nref\bouw{P. Bouwknegt, J. McCarthy, and K. Pilch, ``BRST
Analysis of Physical States for 2D Gravity Coupled to
$c\leq 1$ Matter,'' preprint CERN-TH 6162/91, to appear in
CMP.}%
\refs{\lz{--}\bouw}.
It is possible
that upon elimination of amplitudes involving the non-tachyon
operators one will be left with a set of identities for the
tachyon correlators equivalent to \diffeqs. A second possibility,
suggested by the product of more than two
correlators in \ward, is that there are
singularities in the measure on moduli space%
\foot{or of the differential form $\Theta$ of \witzwie}\
on
high codimension boundaries, and that the simple quadratic
relations of \refs{\witzwie,\erik} must be modified. A third possibility,
of course, is that our relations are distinct from previous results.

\newsec{Conclusions}

We began this paper, and our investigations, with the
motivation of understanding better the time-dependent
backgrounds of string theory.
We have gained a much better understanding of the classical
$S$-matrix of the theory in two dimensions, found an efficient
formalism
for calculation of individual amplitudes,
and obtained some
interesting differential equations for the
partition function $\CF$
on the space of coupling constants $t(p)$.
Solving these equations would lead to a solution
of the sigma model \smbak\ through conformal
perturbation theory. Thus it is interesting to
find an effective procedure for solving these
equations or equivalently the recursion relations
\ward.
Indeed, the (partly experimental) results of
\ref\phasetrn{G. Moore, ``Gravitational Phase
Transitions and the Sine-Gordon Model,''
Yale preprint YCTP-P1-92.}
suggest that some relatively simple and interesting
solutions of \diffeqs\ await discovery.

It is also noteworthy that the collective field equations
are incomplete since they evolve a large class of
smooth initial data to singular field configurations.
This might have some bearing on the issue of classical
singularities in string theory. In particular, the existence
of the free fermion formalism which provides a smooth
description of such ``pathological'' Fermi seas is an
indication that the corresponding singularities in the
Das-Jevicki formalism are not true singularities of the
classical string theory, just of the method of describing
the solution.

\bigskip
\centerline{\bf Acknowledgements}

We are indebted to N. Seiberg for several important
remarks on a previous draft of the paper and for much
correspondence. We also thank S. Ramgoolam
for many discussions and J. Polchinski
for correspondence.
G.M. would also like to thank the Rutgers group for
hospitality and stimulating discussions on this
topic. In particular it is a pleasure to thank T. Banks,
M. Douglas, E. Martinec, S. Shenker,
and A.B. Zamolodchikov.
This work is supported by DOE grant DE-AC02-76ER03075
and by a Presidential Young Investigator Award.

\listrefs
\bye